\documentclass[conference]{IEEEtran}
\IEEEoverridecommandlockouts

\usepackage{algorithmic}
\usepackage{graphicx}
\usepackage{amsmath,amssymb,amsfonts}
\usepackage{cite}
\usepackage{tabulary}
\usepackage{subcaption}
\usepackage{dblfloatfix} 
\usepackage{multirow}
\usepackage{booktabs}
\usepackage{hyperref}
\usepackage{textcomp}
\usepackage{xcolor}
\def\BibTeX{{\rm B\kern-.05em{\sc i\kern-.025em b}\kern-.08em
    T\kern-.1667em\lower.7ex\hbox{E}\kern-.125emX}}
\begin{document}

\title{Trac\'e alternant detector for grading hypoxic-ischemic encephalopathy in neonatal EEG
\thanks{This work is supported by the Wellcome Trust (209325/Z/17/Z). JMOT is supported by Science Foundation Ireland (15/SIRG/3580 and 18/TIDA/6166).}
}

\author{
\IEEEauthorblockN{Sumit A. Raurale$^{1,2}$, Geraldine B. Boylan$^{1,2}$, Sean R. Mathieson$^{1,2}$, William P. Marnane$^{1,3}$, \\ Gordon Lightbody$^{1,3}$, and John M. O'Toole$^{1,2,*}$}
\IEEEauthorblockA{
\textit{$^{1}$Irish Centre for Maternal and Child Health Research (INFANT), University College Cork, Ireland.} \\
\textit{$^{2}$Department of Paediatrics and Child Health, University College Cork,
    Ireland. } \\
\textit{$^{3}$Department of Electrical and Electronic Engineering, University College
    Cork, Ireland.} \\
\textit{$^{*}$jotoole@ucc.ie (corresponding author)}} }

\maketitle

\begin{abstract}
Electroencephalography (EEG) is an important clinical tool to capture sleep-wake cycling. It can also be used for grading injury, known as hypoxic-ischaemic encephalophathy (HIE), caused by lack of oxygen or blood to the brain during birth.
Trac\'e alternant (TA) is a distinctive component of normal quiet sleep which consists of alternating periods of high-voltage activity (bursts) separated by lower-voltage activity (inter-bursts).
This study presents an automated method to grade the severity of injury in HIE, using an automated method to first detect TA activity. 
The TA detector uses the output of an existing method to detect inter-bursts.
Features are extracted from a processed output and then combined in a support vector machine (SVM). 
Next, we develop an HIE grading system using the TA detector by combining different features from the temporal organisation of the detected TA mask, again using an SVM. 
Training and testing for both models use a leave-one-baby-out cross-validation procedure, with model hyper-parameters selected from nested cross validations. 
The TA detector, tested on EEG from 71 healthy term neonates, has an accuracy of 79.1\% (Cohen's $\kappa$=0.55).
The HIE grading system, tested on EEG from 54 term neonates in intensive care, has an accuracy of 81.5\% ($\kappa$=0.74).
These results validate how detecting the presence or absence of TA can be used to quantify the grade of HIE injury in neonatal EEG and open up the possibility of a clinically-meaningful grading system.  

\end{abstract}

\section{Introduction}
\label{introduction}

The electroencephalogram (EEG) provides an effective non-invasive tool to monitor the brain activity of critically ill neonates in the neonatal intensive care unit (NICU) \cite{Scher2002}. Lack of blood and oxygen to the brain around the time of birth can cause permanent brain injury or death. This injury, known as hypoxic-ischaemic encephalopathy (HIE), occurs in approximately 3 to 5 per 1,000 births in high-income countries.
Clinical review of the EEG can correctly identify the degree of HIE severity \cite{murray}.
Around-the-clock interpretation of the EEG is not practical, however, because of the limited resources and specialised expertise required for neonatal EEG. Automated analysis of the EEG could provide continuous monitoring of neonatal brain function and may assist in the critical care of these vulnerable infants.

Neonates spend most of their time sleeping. EEG assessment of the sleep–wake cycle can provide insight into neurological development and brain maturation \cite{Scher2002, Abramsky2020}. An important component of the EEG grading scheme for HIE injury is to assess for the presence of sleep--wake cycling \cite{murray}. In healthy term neonates, 4 major sleep–wake states are evident on the EEG: active sleep, quiet sleep, indeterminate sleep, and wakefulness \cite{Dereymaeker1}. Quiet sleep (QS) can be sub-categorized into 2 states as high-voltage slow-wave ($<$4 Hz) activity and trac\'e alternant (TA) activity \cite{ACNS}. TA activity displays a distinctive pattern of high voltage waveforms known as bursts (typically 50 to 150 $\mu$V peak-to-peak), followed by lower-voltage waveforms known as inter-bursts (typically 25 to 50 $\mu$V peak-to-peak) \cite{ACNS}. Duration of these waveforms range from approximately 2 to 10 seconds.

Although limited in number, there has been recent interest in developing algorithms to detect different sleep states in neonates \cite{Pillay, Ansari, raurale2020ta}. In addition, a limited number of EEG algorithms to grade HIE severity have been proposed \cite{R11, R12, raurale2020cnn, R13, jne}. A subset of these algorithms have focused on detecting clinically relevant attributes of the EEG grade of HIE \cite{R13,matic2015}. Both of these 2 methods extract characteristics of the inter-burst interval in discontinuous activity (which differ to inter-bursts in TA activity) to grade HIE severity \cite{R13,matic2015}.

We aim to continue this approach of clinically-meaningful algorithms by developing a TA detector and using this detector to help grade HIE severity.
Building on our existing work \cite{raurale2020ta}, we develop a machine learning model that detects the presence or absence of TA activity within a neonatal EEG recording. This model is developed using sleep-staging EEG data from 71 healthy term infants.
Next, we extract features of the temporal map of the TA detector, such as the percentage of TA activity, and develop another machine learning model to classify EEG-HIE grades. For this model we use an EEG HIE grading data set of 54 term infants recorded in the NICU. We envisage this model to be 1 part of a larger HIE grading system, consisting of clinically relevant modules with meaningful interpretation.

\begin{figure*}[!t]
	\centering
	\includegraphics[width=0.99\linewidth]{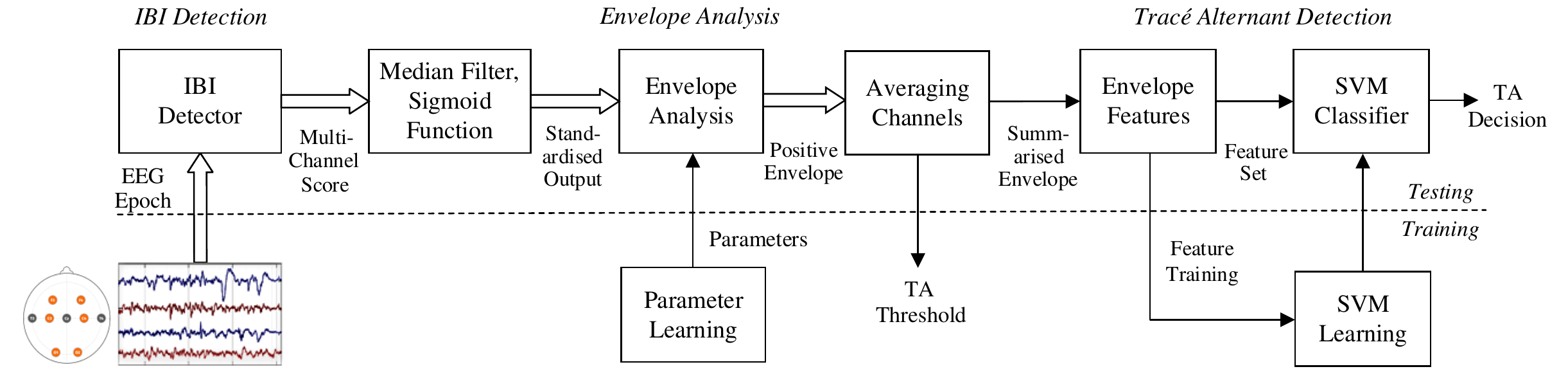}
	\caption{Overview of proposed system for detecting trac\'e alternant (TA) activity. SVM: support vector machine; IBI: inter-burst interval.}
	\label{fig_tasys}
\end{figure*}

\section{Methods}
We first develop an automated TA detector by identifying bursts and inter-bursts within TA activity from the background EEG. The overall structure of the TA detector system is shown in Fig.~\ref{fig_tasys}. Second, we develop a HIE grading system which extracts features from the TA decision function and combines these in a support-vector machine to classify 4 grades of HIE. The grading systems makes a decision from approximately 1 hour of multi-channel EEG.

\subsection{EEG Data and Pre-processing}
Two datasets were used in this study:

\subsubsection{EEG sleep-staging dataset}
EEG was recorded from term infants using a NicoletOne EEG system in Cork University Maternity Hospital, Cork, Ireland with informed and written parental consent \cite{Korotchikova1}. The study was approved by the Clinical Ethics Committee of the Cork Teaching Hospitals. EEG recordings started as soon as possible after birth and continued for up to 1--2 hours to include different sleep states. Five scalp electrodes were used over the frontal and temporal regions with a reference electrode placed between the frontal (Fz) and midline (Cz) region. The EEG was analysed using a 4-channel bipolar montage, derived from these electrodes as F3–T3, F4–T4, T4–Cz and Cz–T3. 
An EEG expert reviewed and annotated instances of TA activity \cite{Korotchikova1}. 
Within the TA activity all bursts and inter-bursts were also annotated.
The dataset consists of a total of 91 EEG recordings, from which a subset of 71 were selected based on the criteria of $>$5 minutes of continuous TA activity. 
EEG data were sampled at 256 Hz during recording and electrode impedance was maintained below 5 k$\Omega$. Movement artefacts, defined as the absolute value of EEG $>$1,500 $\mu$V, were removed. EEG was low-pass filtered to 30 Hz using an finite-impulse response (FIR) filter of length 4,001 samples and then down-sampled to 64 Hz.

\subsubsection{EEG HIE grading dataset}
The EEG was recorded from a NicoletOne EEG system from 54 term infants in the NICU of Cork University Maternity Hospital, Cork, Ireland \cite{R11}. This study was approved by the Clinical Ethics Committee of the Cork Teaching Hospital with written and informed parental consent obtained before EEG recording. To monitor the evolution of the developing encephalopathy, EEG recordings were initiated within 6 hours of birth and continued for up to 72 hours. Nine active electrodes over the frontal, central, temporal, and occipital areas were used for EEG recording. Our analysis used an 8-channel bipolar montage derived from these electrodes as F4–C4, F3–C3, C4–O2, C3–O1, T4–C4, C3–T3, C4–Cz and Cz–C3. To avoid major artefacts, approximately 1-hour EEG epochs were pruned from the continuous EEG. Two EEG experts independently reviewed each epoch and graded according to the system defined by Murray \emph{et al.} \cite{murray}. In cases of disagreement, the experts jointly reviewed the EEG to reach consensus for the final grade. This same dataset has been used by Stevenson \emph{et al}. \cite{R11}, Ahmed \emph{et al}. \cite{R12} and Raurale \emph{et al}. \cite{R13,raurale2020cnn,jne}.

\subsection{Inter-burst Detection}
\begin{figure}[!h]
	\centering
	\includegraphics[width=1.0\linewidth]{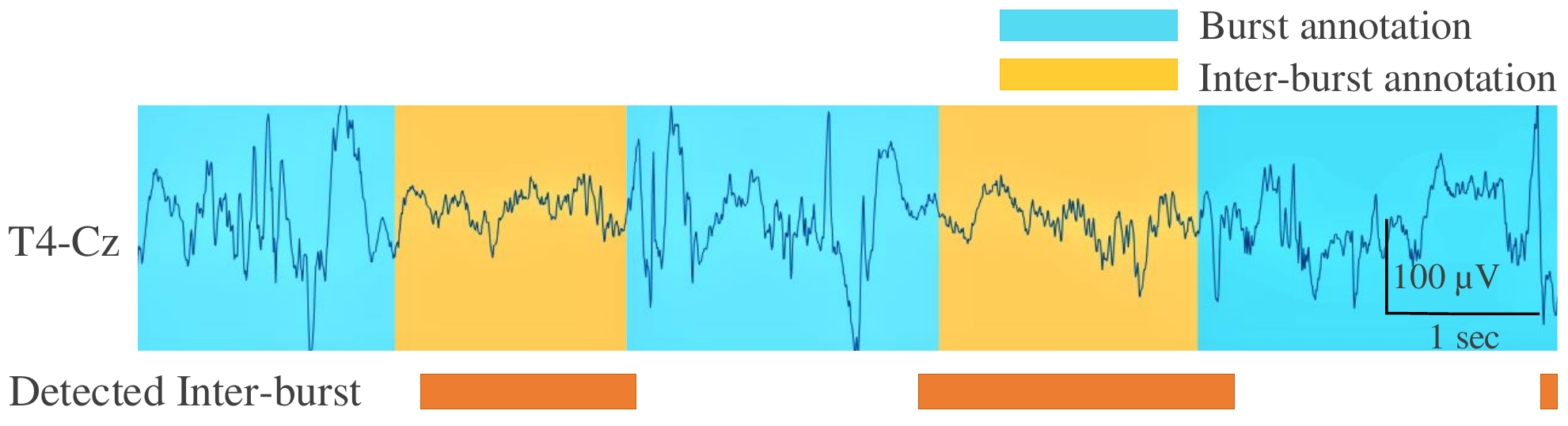}
	\caption{Annotations of bursts and inter-bursts within trac\'e alternant activity and detected inter-bursts for 1 channel EEG.}
	\label{fig_bursup}
\end{figure}

The proposed system architecture uses an existing inter-burst detection method \cite{raurale2020ta} that was developed on the same EEG data.
To classify bursts and inter-bursts present within TA activity, the detector uses multiple features that capture differences in amplitude and spectral shape: envelope, fractal dimension, relative spectral power, measure of spectral fit, and instantaneous frequency across four frequency bands 0.5–4 Hz, 4–7 Hz, 7–14 Hz, and 14–30 Hz; and a frequency-weighted energy measure, the envelope--derivative operator, within the 0.5--10 Hz range \cite{otoole1}.
These features, taken from a preterm inter-burst detection method \cite{otoole1}, were then combined using a linear-kernel support vector machine (SVM). A cross validation procedure, using a leave-one-baby-out, was used in training and testing. Fig.~\ref{fig_bursup} shows an example of 1 channel of EEG with burst and inter-burst annotations and the detector's estimate of inter-bursts. 

\subsection{TA Detection}
We use the output of the inter-burst detector, the confidence score, to build a TA detector which differentiates TA activity from all other EEG states. Although the inter-burst detector was trained and tested only in TA activity, the TA detector is trained and tested on the whole duration of the EEG recording, which includes periods of wakefulness, active sleep, and intermediate sleep.

\begin{figure}[!h]
	\centering
	\includegraphics[width=1.0\linewidth]{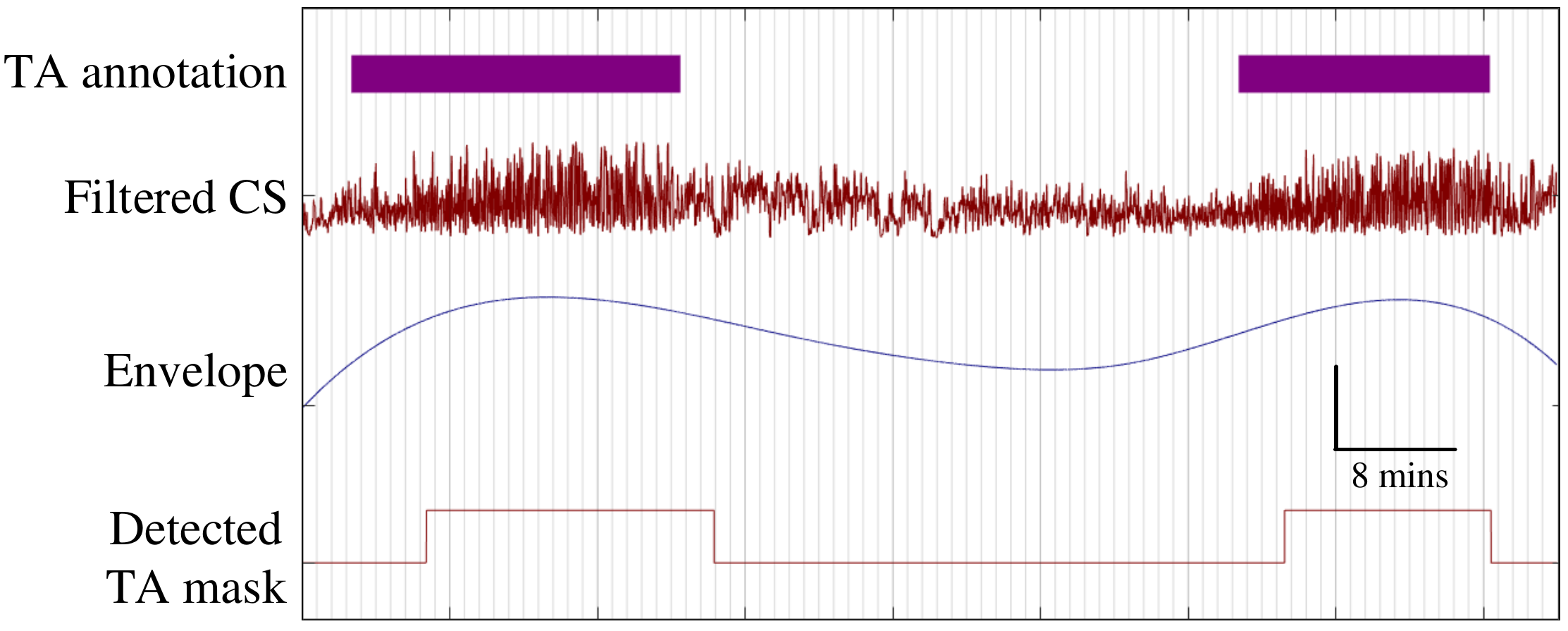}
	\caption{Trac\'e alternant (TA) detection (TA mask), estimated from the smoothed envelope function, which is, in turn, estimated from the the filtered confidence score (CS) of the inter-burst detector. TA annotations (top) from the EEG expert.}
	\label{fig_ta}
\end{figure}

For the first stage of this process, we convert the confidence score of the inter-burst detector, which is an unbounded continuous variable, into a smooth, low-pass envelope function. 
First, we apply a 3-second low-pass moving-median filter to the confidence score to suppress the higher-frequency noise and outliers. The filtered score is then standardised using a sigmoid function to produce a score bounded in the 0 to 1 range. 
Next, an envelope is estimated on the filtered-sigmoid confidence score using local maxima. 
The minimum separation parameter for the local maxima is optimised during training over the range (2.5, 50) seconds with a step size of 2.5-seconds.
Spline interpolation is used to link the peaks, producing a smooth envelope function. A summarised envelope score is produced by averaging across channels. A threshold is used on the summarised envelope function to classify into TA and non-TA activity.
This procedure is similar to what was presented in \cite{raurale2020ta}; this time, we have included the sigmoid function to help scale the envelope function.
Fig.~\ref{fig_ta} shows the visual interpretation of the filtered confidence score output, the smoothed envelope function and the evaluated TA mask. 

In the second stage, the envelope function is used to detect TA activity. The envelope is segmented into 5 minute epochs with 4.5 minute overlap and only those epochs with either full TA activity or full non-TA activity are considered. 
Three features are extracted from the 5-minute envelope epoch: root mean square, median, and maximum value of the envelope.
These features are then combined using a machine learning model. Here, we test 4 different models: a decision tree, a naive Bayes classifier, a linear SVM, and a nonlinear SVM with a radial basis function kernel classifiers.
Training and testing is performed using a leave-one-subject-out (LOSO) cross validation.
Hyperparameters of the radial-basis function SVM (RBF-SVM) were selected from a nested 5-fold cross validation. Values of $\gamma$ and $C$ were selected from the 2D grid of [10$\times$5].
The minimum separation parameter for the envelope function is selected from a grid search within the same (outer) cross validation in a sequential manner rather than a nested cross validation.

For comparison with the machine learning models, a threshold is applied directly to the envelope function to detect TA activity. 

\subsection{HIE Detection}
EEG from the HIE grading dataset was downsampled from 256 Hz to 64 Hz, after application of an anti-aliasing low-pass filter with a cut-off of 30Hz. Each EEG channel was then analysed by the TA detector using a 5-minute segment with a step size of 1 second. 
Two features are extracted from the SVM output of the TA detector, the continuous-valued confidence score (CS). Additionally, 3 features of the binary TA mask are also extracted. These 5 features are as follows:

\vspace{0.10cm}
\noindent 1. CS Median ($CS_{\text{med}}$): central approximate CS value across the 1-hour epoch.

\vspace{0.10cm}
\noindent 2. Coefficient of CS variations ($CS_{\text{CoV}}$): extent of CS variability with respect to the mean CS. Calculated as,
\begin{equation} CS_{\text{CoV}} = \log \left|~CS_{\text{std}} / CS_{\text{mean}}\right| \end{equation}
where $CS_{\text{std}}$ is the standard deviation and $CS_{\text{mean}}$ is the mean of the CS outputs from epoch.


\vspace{0.10cm}
\noindent 3. TA percentage ($TA\%$): overall occurrence of TA activity in the 1-hour EEG epoch. Calculated as the total duration of TA to the overall epoch length,
\begin{equation} TA\% = \frac{100}{L} \sum_{n=0}^{L-1}t[n] \end{equation}
where $t[n]$ is the length-$L$ binary mask. 

\vspace{0.10cm}
\noindent 4. Number of TA instances: counts the number of TA instances over the 1-hour epoch. 

\vspace{0.10cm}
\noindent 5. Maximum duration of TA ($TA_{\text{max}}$): is measure by subtracting the difference between the maximum and the minimum TA length within an epoch as,
\begin{equation} TA_{\text{max}} = \max (t[n]) - \min (t[n]) \end{equation}

\begin{figure}[!h]
	\centering
	\includegraphics[width=0.98\linewidth]{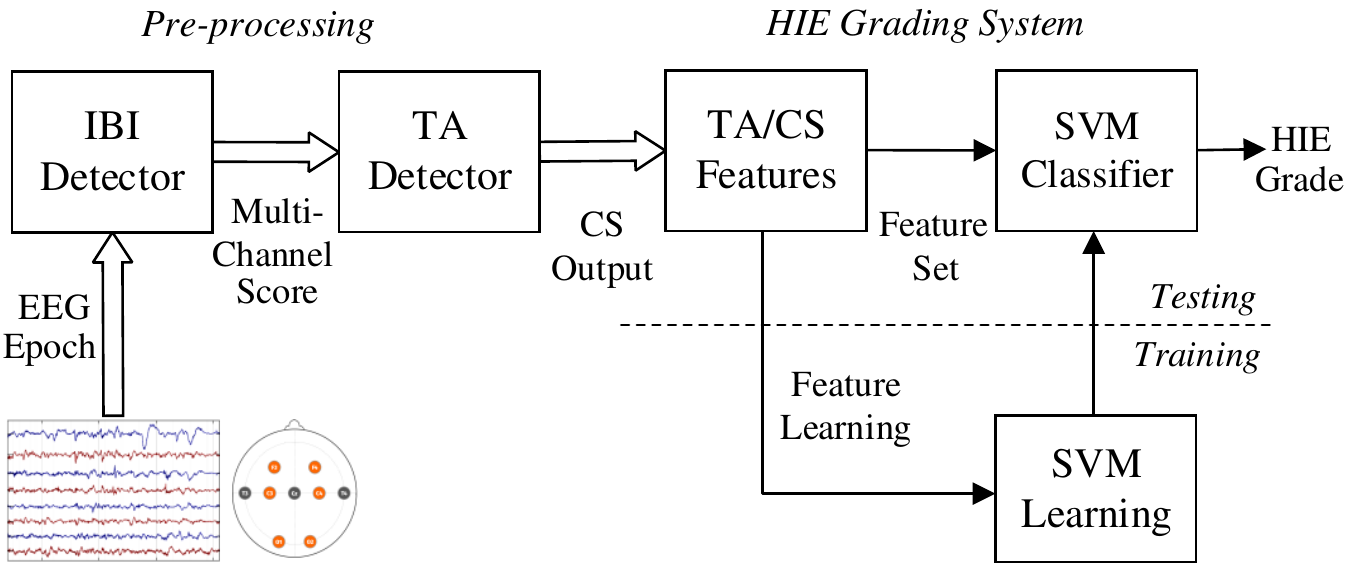}
	\caption{Overview of the proposed system for detecting HIE grade. SVM: support vector machine; TA: trac\'e alternant.}
	\label{hiesystem}
\end{figure}

These features are combined using a RBF-SVM. The SVM classifier is used in one-versus-one configuration to recognise each of the 4 HIE grades. The overall structure of the proposed HIE grading system is shown in Fig.~\ref{hiesystem}.

For training and testing, a LOSO cross-validation method is used. 
The hyper parameters for the RBF-SVM classifier were selected from a 2D grid of [10$\times$5] using a nested 5-fold cross-validation. 

\section{Results}
\label{expresult}

TA detection performance for each of individual envelope features and 3 machine learning models, which combine these features, is in Table~\ref{tafeat_perf}. Performance is assessed using the area under the receiver operator characteristic curve (AUC) using the LOSO cross-validation results.


\begin{table}[!h]
\renewcommand{\arraystretch}{1.0}
\centering
\caption{Performance for individual Envelope features combined with different classifiers for detecting TA and non-TA activity in LOSO cross-validation.}
\label{tafeat_perf}
\resizebox{0.475\textwidth}{!}{
  \begin{tabular}{lcccccc}
    \toprule
    & \multicolumn{3}{c}{Features} & \multicolumn{3}{c}{Classifiers} \\
    \cmidrule(lr){2-4} \cmidrule(lr){5-7}
    & RMS & Max & Median & DT & NB & SVM\\
    \midrule
    AUC & 0.810 & 0.826 & 0.812 & 0.840 & 0.840 & 0.848 \\    
    \bottomrule
\end{tabular}
}
\begin{flushleft} \footnotesize{Key: AUC, area-under receiver operating characteristic curve; DT, Decision tree; NB, Naive Bayes; SVM, support vector machine.} \end{flushleft}
\end{table}





Based on the highest AUC performance, we proceeded with the inclusion of SVM with radial bias function kernel classifier for TA detection.
Table~\ref{taperf} shows the performance metrics for testing the 5-minute EEG segment for classifying TA or non-TA activity using the LOSO cross-validation. 
Confidence intervals are generated from 1,000 bootstrap iterations on a per-baby, and not epoch, basis.
The summarised envelope function is also included for comparison. We used a threshold value of 0.93 for the envelope function as this gives approximately equal sensitivity and specificity (69.9\% and 69.9\%).

\begin{table}[!h]
\renewcommand{\arraystretch}{1.25}
\centering
\caption{Trac\'e alternant detection performance comparing the envelope function with the machine learning approach using a support vector machine (SVM).}
\label{taperf}
\setlength\tabcolsep{3.25pt}
\resizebox{0.49\textwidth}{!}{
\begin{tabular}{lcccc}
\toprule
& \multicolumn{2}{c}{Envelope} & \multicolumn{2}{c}{SVM} \\
\cmidrule(lr){2-3} \cmidrule(lr){4-5}
& Median & (95\% CI) & Median & (95\% CI)  \\
\midrule
AUC & 0.774 & (0.737--0.811) & 0.848 & (0.805--0.888) \\
kappa & 0.420 & (0.347--0.485) & 0.549 &  (0.463--0.627) \\
Accuracy (\%) & 73.1  & (69.9--76.1) & 79.1 &  (75.2--82.7)\\
F1-score (\%) & 62.7  & (56.9--67.8) & 71.5 &  (65.1--77.2)\\
\bottomrule
\end{tabular}
}
\begin{flushleft} \footnotesize{Key: CI, confidence intervals.} \end{flushleft}
\end{table}

\begin{figure}[h!]
\centering
\includegraphics[width=0.83\linewidth]{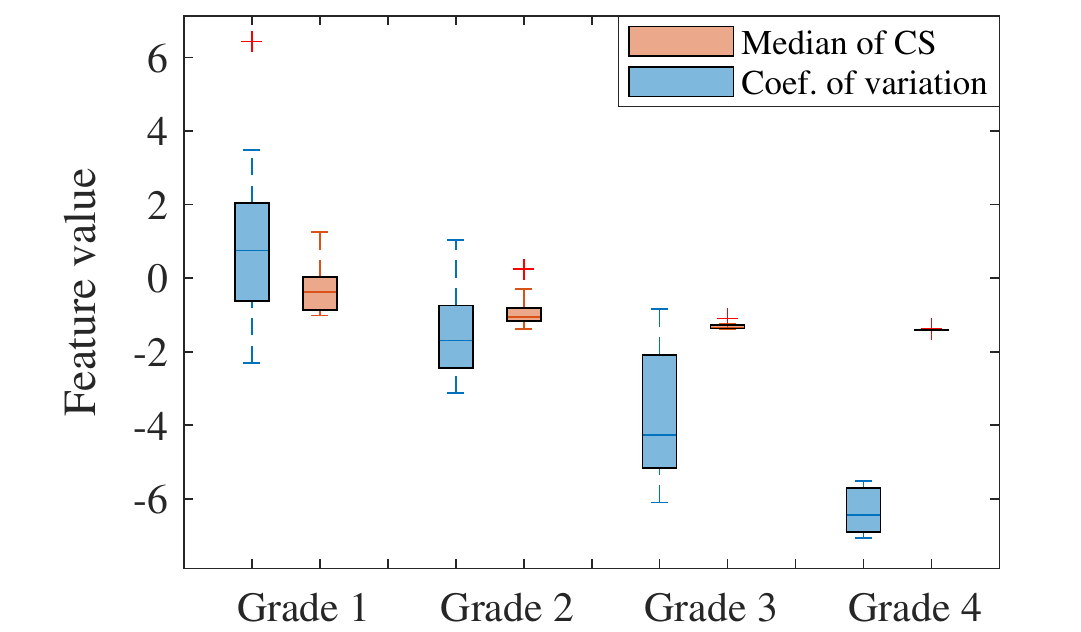}
\includegraphics[width=0.83\linewidth]{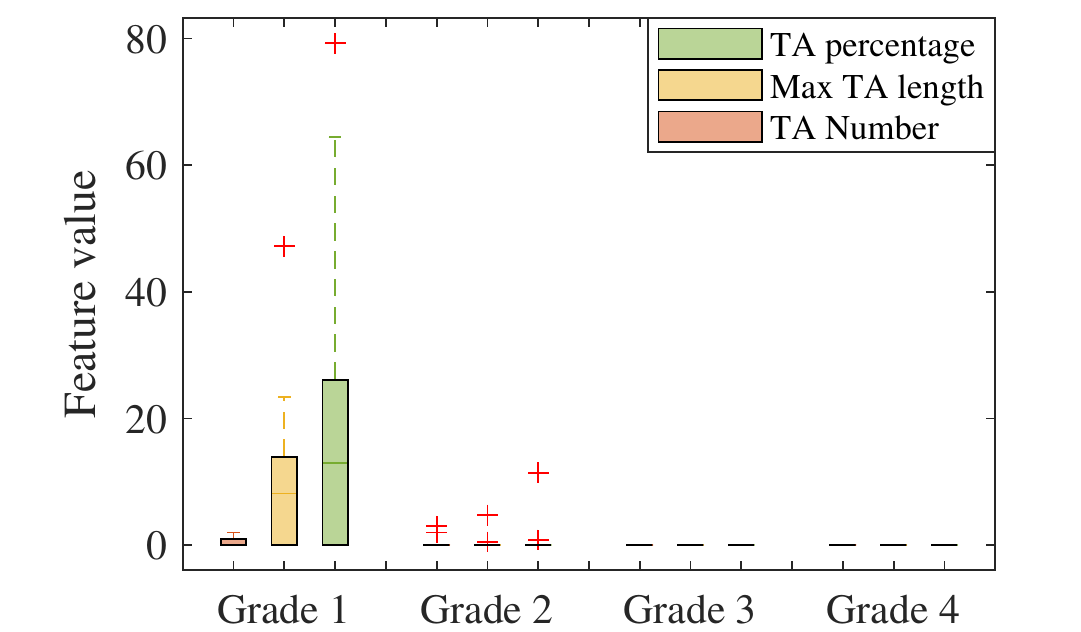}
\caption{Distribution of 5 features estimated from the trac\'e alternant (TA) detector for each EEG grade. 
Top: features of the continuous confidence score (CS); coefficient of variation is log-transformed.
Bottom: features of the binary output from the TA detector.
Maximum TA length is measured in minutes; TA percentage is represented as a percentage. Box represent inter-quartile range, dotted lines represent median values, and error bars represent 95th centiles.
}
\label{tafeatperf}
\vspace{-0.3cm}
\end{figure}

The EEG--HIE grading model is developed using the RBF--SVM TA detector, trained from the first LOSO iteration of the sleep EEG dataset. Fig.~\ref{tafeatperf} illustrates the features extracted from the TA detector over 1-hour EEG epochs for each of the 4 HIE grades. 

The 2 features of CS, median and log-absolute coefficient of variation, show changes from positive to negative values as the HIE grade increases. Also, the number of TA instances, the overall percentage of TA, and the maximum length of TA clearly reduce as the grade of HIE increases. We find clear degrees of separation between the grade 1 and grade 2 for all 3 binary-TA features. The estimated TA number in grade 1 EEG epoch varying from 0 to 3 number, 0 to 2 in grade 2 and nearly 0 for grade 3 and 4 EEGs. The maximum estimated TA length ranges from 1 to 24 minutes for grade 1, under 5 minutes for grade 2 and falls to 0 minute for grade 3 and grade 4. The percentage of TA activity ranges up to 64\% for grade 1 and up to 13\% for grade 2. The TA detector finds no TA activity in grade 3 and grade 4. Thus, the binary-TA features show separation for grade 1 and grade 2 while the CS features shows separation for all grades, including grade 3 and 4.

Table~\ref{confusion} presents the confusion matrix for the LOSO cross-validation using the 5 features of the TA detector across the 1-hour EEG epoch. It shows that 44/54 EEGs were correctly classified, resulting in an accuracy of 81.5\% ($\kappa$=0.74). Most misclassification occurred between grade 3 and 4, which from Fig.~\ref{tafeatperf} appears to be the harder grades to separate using TA features alone.

\begin{table}[!h]
\renewcommand{\arraystretch}{1.30}
\centering
\vspace{0.11cm}
\caption{Confusion matrix based on the HIE system output and actual grades.}
\label{confusion}
\setlength\tabcolsep{2.9pt}
\begin{tabular}{c l l l l c r}
\midrule 
Actual & \multicolumn{4}{c}{System's Output} & \multirow{2}{*}{~Total~} & \multirow{2}{*}{~False}\\
 \cline{2-5}
grade & ~~~1 & ~2 & ~3 & ~4 & & \\ 
\midrule
1 & {\bf 17~(77.3\%)} & ~5 & ~0 & ~0 & 22 & 5~~~\\
2 & ~~~2 & {\bf 12~(85.7\%)} & ~0 & ~0 & 14 & 2~~~\\
3 & ~~~0 & ~2 & {\bf 9~(75.0\%)} & ~1 & 12 & 3~~~\\
4 & ~~~0 & ~0 & ~0 & {\bf 6~(100\%)} & 6 & 0~~~\\
\midrule
Total~ & ~~19 & 19 & 9 & ~7 & 54 & 10~~~\\
\midrule
\end{tabular}
\vspace{-0.3cm}
\end{table}

The performance of other state-of-the-art systems which use the same EEG--HIE dataset compared with our proposed system are presented in Table~\ref{compare}. The method from Stevenson \emph{et al}. (TFDfeat) \cite{R11} extracted a complex feature set from the instantaneous frequency and the instantaneous amplitude measures of a quadratic time-frequency distribution, combined using a linear classifier. Raurale \emph{et al}. (CNN1d) \cite{raurale2020cnn} extracts convolutional features from raw EEG using data-driven deep learning approach. Ahmed \emph{at al}. (GSVfeat) \cite{R12} used 55 generic features combined using super-vectors from a Gaussian mixture model and classified in a RBF-SVM. Raurale \emph{et al}. (IBIfeat) \cite{R13} used two features of inter-burst activity evaluated from a burst detector and combined in a multi-layer perceptron. Also, Raurale \emph{et al}. (TFDCNN) \cite{jne} extracts 2-dimension time-frequency distribution map from raw EEG to extract time, frequency and time-frequency convolutional features as data-driven deep learning approach. In contrast, our proposed system operates on evaluating only TA activity, yet it is still better performance to the TFDfeat \cite{R11} and IBIfeat \cite{R13} systems and equal performance to the CNN1d \cite{raurale2020cnn} system.  

\begin{table}[h]
\renewcommand{\arraystretch}{1.5}
\centering
\caption{Comparison of the proposed EEG grading method with existing techniques on the same database.}
\label{compare}
\setlength\tabcolsep{3.0pt}
\begin{tabular}{r | c c c c c c}
\toprule
\textbf{HIE Grading} & TFDfeat & GSVfeat & CNN1d & IBIfeat & TFDCNN & Proposed \\
\textbf{Method} & \cite{R11} & \cite{R12} & \cite{raurale2020cnn} & \cite{R13} & \cite{jne} & method \\
\midrule
\textbf{Accuracy} & 77.8\% & 87.0\% & 81.5\% & 77.8\% & 88.9\% & 81.5\%\\ 
\bottomrule
\end{tabular}
\end{table}

\section{Discussion and Conclusions}
\label{discussion}

We present a novel approach for detecting TA activity in the EEG of term neonates by combining features of an envelope function generated from the confidence score of an inter-burst detector. The detector is developed on an EEG data set of 71 healthy term neonates; testing results indicate moderate-to-good performance (AUC = 0.85, $\kappa=0.55$, accuracy = 79\%, F1 = 72\%). 
This TA detector is then applied to an EEG data set of 54 term infants in the NICU.
Features of the confidence score and the binary mask of the TA detector are then used to develop a HIE grading system. This system achieves comparable accuracy (82\%) with other complex, state-of-the-art methods. 

Sleep--wake cycling is an important tool for assessing recovery in term infants with moderate to severe HIE \cite{murray}. We focused on identifying the presence TA, an essential and distinctive component of the cycle. 
We expect little to no sleep cycling in grades 2 to 4 of HIE. The TA detector did detect some TA activity in grade 2 (see Fig.~\ref{tafeatperf}). This may be due to incorrectly detected periods of discontinuity by the IBI detector, resulting in false TA detection. For grades 3 and 4, no TA activity was detected. In contrast, the 2 features of the continuous-valued confidence score, also in Fig.~\ref{tafeatperf}, do indicate a separation for grades 3 and 4. It is likely that these features provide the HIE SVM model with sufficient information to classify grades 3 and 4, as shown in Table~\ref{confusion}.  


In conclusion, we validate the use of an EEG TA detection method to classify HIE grades. Future systems will combine interpretable sub-systems such as this one with others, such as discontinuity detectors for example \cite{R13}. 
By combining meaningful sub-systems, we aim to improve both performance and clinical utility.


\end{document}